# Enriching Bibliographic Data by Combining String Matching and the Wikidata Knowledge Graph to Improve the Measurement of International Research Collaboration

Ba Xuan Nguyen[1], Jesse David Dinneen[1] and Markus Luczak-Roesch[1]

[1] *ba.nguyen @vuw.ac.nz, jesse.dinneen@vuw.ac.nz, markus.luczak-roesch@vuw.ac.nz*
School of Information Management, Victoria University of Wellington, Wellington (New Zealand)

## Introduction

As publications with international research collaborations receive, on average, a higher number of citations (Glänzel et al., 2001), researchers are incentivised to collaborate. As a result, both the number and the ratio of international co-authored papers have risen (Fortunato, 2018). Research managers and policy makers are interested in measuring international research collaboration (IRC; Luukkonen, 1993), for example to determine if relevant policies are effective.

*Problem statement*
Currently, the common measure of IRC is a count of co-authored research publications (Chen, Zhan, & Fu, 2018) reported in various data sets (e.g., Web of Science, Google Scholar). The most commonly used data sets in IRC research are SCI/Web of Science and Scopus (Guerrero Bote et al., 2013; Luukkonen, 1993). Each set entails considerable practical challenges for researchers; for example, only 500 records can be downloaded from WoS at a time, or 2,000 from Scopus. Further, these sets (and Google Scholar) do not have as comprehensive general coverage as, for example, Microsoft Academic Graph (MAG; Paszcza, 2016; Sinha et al., 2015), and may not have as complete domain-specific coverage as, for example, ACM Digital Library (ACM DL) and IEEE Xplore provide for computer science. All sources have in common one considerable practical challenge, however: measuring IRC requires mapping the affiliation data from each publication to the relevant countries, and no method for doing this has been previously (e.g., in prior work). The task is non-trivial because, for example, there are many records with varying affiliation formats (e.g., ending with country, or with state/province, or just an institution like "McGill University") or dirty data (e.g., ending in "#TAB#"), and there is no standard method for associating such values with the parent country. In short, measuring IRC is desirable, but currently difficult. Here we describe a method to address this difficulty, and evaluate it using both general and domain-specific data sets.

## Preparation of data sets

In this paper we test our method on MAG, a general scholarly bibliographic data set, and ACM DL, a scholarly bibliographic data set containing works published by the Association for Computing Machinery and primarily related to computer science. To make the results of our evaluations comparable across the two sets we filtered out records from MAG that were not relevant to computer science: a list of fields of study (FOS) was compiled from records present in both ACM and MAG, and the 38 top FOS terms (94% of papers in the overlap) were used to filter out irrelevant works. Overlapping papers were also filtered from the ACM set to make the sets distinct. Finally, single-author records were filtered out to identify only co-authored papers. Table 1 summarises the results.

**Table 1. Summary of data sets used**

| Features | ACM DL | MAG |
|---|---|---|
| Total works | 182,791 | 212,689,976 |
| Date range | 1951-2017 | 1965-2017 |
| Unique, co-authored, CS works | 121,672 | 594,036 |

## Contributed method

Two steps were implemented to identify the collaborating countries using the authors' affiliation data in co-authored papers. First, for records with author affiliation data (i.e., in the *org* field in MAG and *affiliation* in ACM) containing names or abbreviations of countries or their component parts (e.g., US states), substrings of the location names were extracted and matched to a list of countries. The UK was considered in this study as a whole entity for all its component parts. Second, for records having no country names or state information, we then used the remaining information (e.g., university name) to query the SPARQL endpoint of Wikidata[1] executing the following query[2]:

```
PREFIX schema: <http://schema.org/>
PREFIX wdt:
<http://www.wikidata.org/prop/direct/>
SELECT ?countryLabel WHERE
{<https://en.wikipedia.org/wiki/[AFFILIATION]>
schema:about ?datalink. ?datalink wdt:P17
?country.SERVICE wikibase:label
{bd:serviceParam wikibase:language "en".}}
```

This query returns English names of countries associated with the location data if there is a matching Wikidata item. For example, querying "McGill University" returns Canada. We implemented both steps in *R* and have made the source code freely available for use in future work.[3]

**Method evaluation**

Our method identifies countries for approximately 70%-80% records in each data set (details in Table 2), with the remaining records being either unidentified (~15%) or unidentifiable (8-14%) because of empty affiliation values (e.g., NA). Specifically, while the substring matching approach identifies countries for 60-70% of records, the Wikidata querying approach adds an additional 11% in ACM and 8.41% in MAG. In other words, the method provided identifies approximately 85% of the possible records. These results suggest our approach succeeds in matching the majority of bibliographic records, in both general and domain-specific data sets, to the relevant countries.

**Table 2. Results of country identification**

| Results | ACM DL | MAG |
|---|---|---|
| Affiliations | 384,672 | 831,888 |
| NA, Null, etc values | 52,454 (13.66%) | 65,674 (7.89%) |
| Country names identified | 136,671 (35.60%) | 549,992 (66.11%) |
| Component parts identified | 98,622 (25.69%) | 31,738 (3.82%) |
| Identified by Wikidata | 42,050 (10.94%) | 69,985 (8.41%) |
| Not identified (Other values) | 54,875 (14.29%) | 116,982 (14.06%) |

---

[1] https://query.wikidata.org/sparql
[2] "[AFFILIATION]" in this query gets replaced with the remaining information extracted

**Conclusion**

A current problem in IRC research is that it is difficult to identify countries by the affiliation information that bibliographic records provide. Previously, no method was available to overcome this, so methods were *ad hoc,* impractical, and likely inconsistent with each other, potentially resulting in varying results across even studies using the same data sets, or worse, preventing IRC measurement altogether. Here we provided and evaluated a novel method for addressing the problem, using substring matching and the SPARQL endpoint of the Wikidata knowledge graph. The method appears promising for use with other data sets as well, especially given that Wikidata will continue to grow and thus improve in matching affiliations to countries.

---

[3] https://github.com/baxuan/IRCM/tree/master/Comparing-data-sets